\documentclass{pasj00}
\usepackage[round]{natbib}

\begin{document}
\SetRunningHead{M. Serino et al.}
{Peculiarly Narrow SED of GRB\,090926B}
\Received{}

\title{Peculiarly Narrow SED of GRB\,090926B with MAXI and Fermi/GBM}

\author{
Motoko \textsc{Serino},\altaffilmark{1}
Atsumasa \textsc{Yoshida},\altaffilmark{2}
Nobuyuki \textsc{Kawai},\altaffilmark{3}
Yujin E. \textsc{Nakagawa},\altaffilmark{4}
Tatehiro \textsc{Mihara},\altaffilmark{1}
Yoshihiro \textsc{Ueda}\altaffilmark{5}
Satoshi \textsc{Nakahira},\altaffilmark{1}
Satoshi \textsc{Eguchi},\altaffilmark{5}
Kazuo \textsc{Hiroi},\altaffilmark{5}
Masaki \textsc{Ishikawa},\altaffilmark{6}
Naoki \textsc{Isobe},\altaffilmark{7}
Masashi \textsc{Kimura},\altaffilmark{8}
Hiroki \textsc{Kitayama},\altaffilmark{8}
Mitsuhiro \textsc{Kohama},\altaffilmark{9}
Takanori \textsc{Matsumura},\altaffilmark{10}
Masaru \textsc{Matsuoka},\altaffilmark{1,9}
Mikio \textsc{Morii},\altaffilmark{3}
Motoki \textsc{Nakajima},\altaffilmark{11}
Hitoshi \textsc{Negoro},\altaffilmark{12}
Megumi \textsc{Shidatsu},\altaffilmark{5}
Tetsuya \textsc{Sootome},\altaffilmark{1}
Kousuke \textsc{Sugimori},\altaffilmark{3}
Mutsumi \textsc{Sugizaki},\altaffilmark{1}
Fumitoshi \textsc{Suwa},\altaffilmark{12}
Takahiro \textsc{Toizumi},\altaffilmark{3}
Hiroshi \textsc{Tomida},\altaffilmark{9}
Yohko \textsc{Tsuboi},\altaffilmark{10}
Hiroshi \textsc{Tsunemi},\altaffilmark{8}
Shiro \textsc{Ueno},\altaffilmark{9}
Ryuichi \textsc{Usui},\altaffilmark{3}
Takayuki \textsc{Yamamoto},\altaffilmark{1}
Kazutaka \textsc{Yamaoka},\altaffilmark{2}
Makoto \textsc{Yamauchi},\altaffilmark{13}
Kyohei \textsc{Yamazaki},\altaffilmark{10}
and the MAXI team} 

\altaffiltext{1}{MAXI team, Institute of Physical and Chemical Research (RIKEN), 2-1 Hirosawa, Wako, Saitama 351-0198}
\email{motoko@crab.riken.jp}
\altaffiltext{2}{Department of Physics and Mathematics, Aoyama Gakuin University,\\ 5-10-1 Fuchinobe, Chuo-ku, Sagamihara, Kanagawa 252-5258}
\altaffiltext{3}{Department of Physics, Tokyo Institute of Technology, 2-12-1 Ookayama, Meguro-ku, Tokyo 152-8551}
\altaffiltext{4}{Research Institute for Science and Engineering, Waseda University, 17 Kikui-cho, Shinjuku-ku, Tokyo 162-0044}
\altaffiltext{5}{Department of Astronomy, Kyoto University, Oiwake-cho, Sakyo-ku, Kyoto 606-8502}
\altaffiltext{6}{School of Physical Science, Space and Astronautical Science, The graduate University for Advanced Studies (Sokendai), Yoshinodai 3-1-1, Chuo-ku, Sagamihara, Kanagawa 252-5210}
\altaffiltext{7}{Institute of Space and Astronautical Science(ISAS), Japan Aerospace Exploration Agency(JAXA) ,3-1-1 Yoshino-dai, Chuo-ku, Sagamihara, Kanagawa 252-5210}
\altaffiltext{8}{Department of Earth and Space Science, Osaka University, 1-1 Machikaneyama, Toyonaka, Osaka 560-0043}
\altaffiltext{9}{ISS Science Project Office, Institute of Space and Astronautical Science (ISAS), Japan Aerospace Exploration Agency (JAXA), 2-1-1 Sengen, Tsukuba, Ibaraki 305-8505}
\altaffiltext{10}{Department of Physics, Chuo University, 1-13-27 Kasuga, Bunkyo-ku, Tokyo 112-8551}
\altaffiltext{11}{School of Dentistry at Matsudo, Nihon University, 2-870-1 Sakaecho-nishi, Matsudo, Chiba 101-8308}
\altaffiltext{12}{Department of Physics, Nihon University, 1-8-14 Kanda-Surugadai, Chiyoda-ku, Tokyo 101-8308}
\altaffiltext{13}{Department of Applied Physics, University of Miyazaki, 1-1 Gakuen Kibanadai-nishi, Miyazaki, Miyazaki 889-2192}

\KeyWords{gamma rays: bursts --- methods: data analysis} 

\maketitle

\begin{abstract}

  The monitor of all-sky X-ray image (MAXI) Gas Slit Camera (GSC) on 
 the International Space Station (ISS) detected a gamma-ray burst (GRB)
 on 2009, September 26, GRB\,090926B. This GRB had extremely hard spectra 
 in the X-ray energy range.  
 Joint spectral fitting with the Gamma-ray Burst Monitor on the Fermi 
 Gamma-ray Space Telescope shows that this burst has peculiarly narrow 
 spectral energy distribution and is represented by Comptonized blackbody 
 model.  This spectrum can be interpreted as photospheric emission
 from the low baryon-load GRB fireball.  Calculating the parameter of 
 fireball, we found the size of the base of the flow
 $r_0 = (4.3 \pm 0.9) \times 10^{9} \, Y^{\prime \, -3/2}$ cm and Lorentz factor of 
 the plasma $\Gamma = (110 \pm 10) \, Y^{\prime \, 1/4}$, where $Y^{\prime}$ is a ratio 
 between the total fireball energy and the energy 
 in the blackbody component of the gamma-ray emission.
 This $r_0$ is factor of a 
 few larger, and the Lorentz factor of 110 is smaller by also factor of a 
 few than other bursts that have blackbody components in the spectra.  
\end{abstract}

\section{Introduction}
 A discovery of the afterglows of gamma-ray bursts (GRBs) made it clear 
 that GRBs are at cosmological distance and emit enormous energy.  The
 spectra of GRB prompt emission have been expressed empirically
 with a smoothly broken power-law model (``Band Function'')
 \citep{1993ApJ...413..281B}. In order to explain the huge energy 
 and non-thermal spectral shape, models of synchrotron emission from 
 shock-accelerated electrons in 
 the relativistic outflow \citep[e.g.][]{1994ApJ...430L..93R} 
 are suggested.
 However, sometimes the low-energy component is very hard and the 
 power-law photon index $\alpha$
 \footnote{We use the photon index $\alpha$ in the context of $E^{\alpha}$, where $E$ is the energy, throughout this paper.}
 becomes greater than the theoretical limit of $2/3$ \citep{2002ApJ...581.1248P}.
 This observational fact has been explained in many ways such as
 non-thermal \citep[e.g.][and references therein]{2009ApJ...702L..91M, 2010ApJ...725.1137L} 
 and thermal \citep[e.g.][]{2000ApJ...530..292M, 2004ApJ...614..827R, 2010MNRAS.407.1033B} models, but 
 reasons for such hard spectra have not been understood clearly.  
 For example, 
 \citet{2003A&A...406..879G} suggest that some thermal models are 
 consistent with observed spectral characteristics of several GRBs
 with extremely hard spectra. On the other hand, 
 \citet{2005PASJ...57.1031S} remark that ``jitter'' radiation 
 \citep[e.g.][]{1973ApJ...183..611E, 2000ApJ...540..704M} is
 one of possible mechanisms of reproducing spectral indices $\alpha>-2/3$.

 A GRB was triggered by Swift/BAT and Fermi/GBM at 21:55 on 2009
 September 26. 
 The quick results of this burst, GRB\,090926B, are reported 
 to GCN by the Swift team \citep{2009GCNR..246....1G} and the Fermi team
 \citep{2009GCN..9957....1B}. Both of them conclude that the burst has 
 a hard spectrum, whose photon index is $\alpha > -2/3$, 
 above the critical value of synchrotron shock models, so-called 
 ``line of death''.  GRB\,090926B was also observed with MAXI/GSC.  
 MAXI/GSC can examine a low energy portion of the spectra of GRBs below 10 keV, 
 while Swift/BAT or Fermi/GBM observes $\gtrsim$ 10 keV.  The observational 
 results below 10 keV may give even more severe constraints to 
 interpretation of the spectra.
 In this paper, we report the observational results obtained with MAXI/GSC
 on GRB\,090926B, and discuss their interpretation.

\section{MAXI Observations and Data Analysis}
 \subsection{Observations and Data reduction}
 Monitor of All-sky X-ray Image (MAXI) is a mission mounted on the 
 International Space Station (ISS) \citep{2009PASJ...61..999M}.  The 
 cameras of MAXI scan X-ray sources as the ISS rotates around the Earth.  
 MAXI has two scientific instruments: the Gas Slit Camera 
 \citep[GSC;][]{mihara.gsc} and the Solid-state Slit Camera 
 \citep[SSC;][]{2011PASJ...63..397T}.  Since GRB\,090926B was out of the SSC 
 field of view, only the data of the GSC were available for the burst.

 The GSC consists of twelve one-dimensional position sensitive 
 proportional counters sensitive to 2--30 keV photons.
 Six GSC counters constitute an instantaneous field of view (FOV) of
 3$^{\circ}\times$160$^{\circ}$ pointing toward the ISS motion
 (GSC-H), and the other six counters another FOV 
 pointing to the zenith (GSC-Z). 
 GSC-H and GSC-Z work together and covers 85\% of the whole sky every 
 ISS orbit. The transit time of a camera for a source is about 40--150 
 s, depending on the incident angle to the camera
 \citep[See ][for details]{2011arXiv1102.0891S}.

\begin{figure}[htbp]
  \begin{center}
   \rotatebox{0}{
    \FigureFile(80mm,60mm){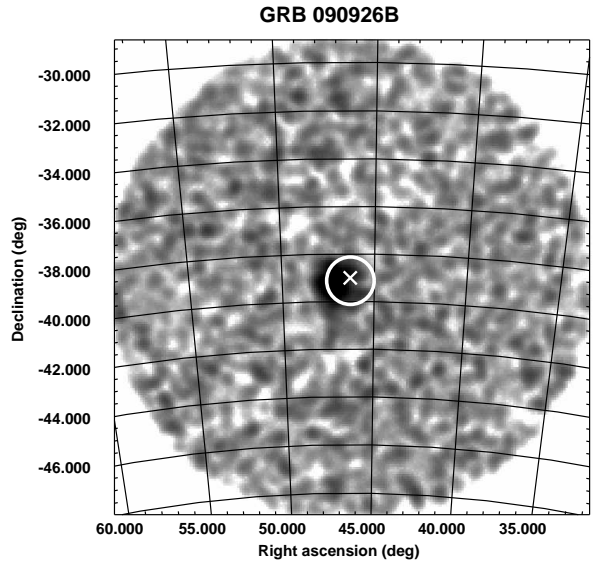}
    }
  \end{center}
  \caption{MAXI GSC image of GRB\,090926B. The MAXI error circle reported 
  to GCN \citep{2009GCN..9943....1M} is shown. The ``X'' mark denotes the 
  position of the GRB derived by Swift XRT \citep{2009GCNR..246....1G}. }
 \label{fig:im}
\end{figure}

 GRB\,090926B, the second GRB detected with MAXI, was observed with
 three cameras of the GSC: camera 0, 1, and 7. Total effective area for these 
 three cameras is about 14 cm$^2$ at maximum%
 \footnote{The effective area of MAXI/GSC to a source changes during the 
 $\sim 40$ s transit time as a triangular curve.}.
 Figure~\ref{fig:im} is an image of GRB\,090926B observed with the GSC. 
 The GSC scanned the field from the right to the left in the image.
 The position localized by the GSC is reported to GCN by 
 \citet{2009GCN..9943....1M} with an error circle of a radius of about 
 1 degree, which is shown in Figure~\ref{fig:im}.  
 The position of the X-ray afterglow observed by Swift/XRT is pointed with
 the ``X'' mark in the same figure.  The bright region of the image is clearly 
 shifted toward the left relative to the XRT position.  This is mostly
 because the intensity has changed during the scan.
 For a steady source, a point spread function must be symmetric 
 on the source. In the case of GRB\,090926B, the burst started when the
 source came across the FOV by 1/4, then brightened after 
 the source passed the center of the FOV.

 We also look into the data of other transits $\sim$ 5500 s before and 
 after the burst. We cannot find any emission down to the flux
 limit of about $3 \times 10^{-10}$ erg cm$^{-2}$ s$^{-1}$ (3$\sigma$)
 in the 4--10 keV band for either scan. 
 According to the swift observation of the X-ray 
 afterglow \citep{2009GCNR..246....1G}, the energy flux at the time of
 the earliest MAXI
 scan after the burst was $7 \times 10^{-12}$ erg cm$^{-2}$ s$^{-1}$ (0.3--10 keV), 
 which was below the MAXI's detection limit.

 For the light curve and spectral analyses,
 we use X-ray event data with processing version 0.3 provided by the MAXI team.
 This data set has time resolution of $\le$ 
 1 ms and 1200 PI channel (1 PI = 0.05 keV).  In addition to the GSC data,
 we use Fermi/GBM archival data%
 \footnote{http://fermi.gsfc.nasa.gov/ssc/data/access/gbm/},
 in order to compensate the limited energy range of the GSC.
 We use {\it XSELECT} ver. 2.4a and {\it XSPEC} ver. 12.5.0ac for
 the data selection and spectral analyses, respectively.

 \subsection{Light curves}
\begin{figure}[htbp]
  \begin{center}
   \rotatebox{0}{
    \FigureFile(80mm,60mm){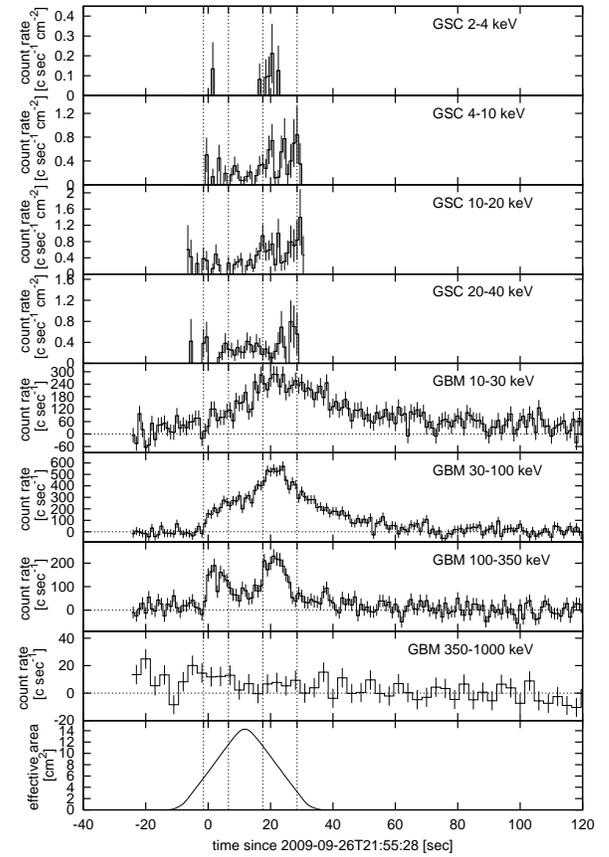}
    }
  \end{center}
  \caption{The light curves of GRB\,090926B observed with MAXI/GSC and 
  Fermi/GBM. The light curves of 
  GSC are corrected for the effective area.  The change
  of the effective area is shown in the bottom panel. The vertical
  dashed lines indicate the borders of the time intervals for 
  spectral analyses.
  }
  \label{fig:lc}
\end{figure}

 Figure~\ref{fig:lc} shows the light curves of GRB\,090926B with
 MAXI/GSC and Fermi/GBM. The count rate of MAXI/GSC is corrected for
 the effective area, which is shown in the bottom panel of the figure.
 Although MAXI/GSC covered only the first 30 s of the prompt emission, the 
 burst actually lasted for more than 50 s; \citet{2009GCNR..246....1G}
 (Swift) and \citet{2009GCN..9957....1B} (Fermi) report the burst
 duration of $T_{90}=109.7\pm11.3$ s and $T_{90}=81\pm13$ s,
 respectively. As seen in Figure~\ref{fig:lc}, it shows a relatively
 hard spectrum in the first 15 s. By contrast, the spectrum becomes relatively
 soft in the following part of the burst.

 \subsection{Spectral analysis}

 The results of the time averaged spectra of the burst observed with
 Swift and Fermi are reported to GCN by \citet{2009GCNR..246....1G}
 and \citet{2009GCN..9957....1B}, respectively. Both teams mention
 that the spectrum can be fit with a cutoff power law model.
 \citet{2009GCN..9957....1B} integrate the spectrum in the first 48.6
 s, and find the peak energy in the $E F_E$ spectrum (where $F_E$ is
 the energy flux at energy $E$) of $E_{\mathrm peak} = 91\pm2$ keV
 and $\alpha$ = -0.13 $\pm$ 0.06. \citet{2009GCNR..246....1G} adopt a
 longer (154.8 s) integration time and find $E_{\mathrm peak}$ = 78.3
 $\pm$ 7.0 keV and $\alpha$ = -0.52 $\pm$ 0.24. These photon indices
 are remarkably large, well above the line of death.
 
 Since these results are derived from analyses above $\sim$ 10 keV, we
 examine whether or not this power-law is extended to energies below
 10 keV, using the MAXI/GSC data. First, we extract a time averaged
 spectrum from $T_0-$1.5 to $T_0+$28.5 s, where $T_0$ is the trigger
 time of Fermi (2009 September 26, 21:55:28). To reduce the
 statistical uncertainty of the photon index due to the limited energy
 range of the GSC, we then simultaneously fit the spectra of the GSC
 together with that of Fermi/GBM.  We test both a cutoff power law
 model and the empirical ``GRB model'' \citep{1993ApJ...413..281B},
 which has 4 free parameters, a photon index in the lower energy band
 $\alpha$, that in the higher band $\beta$, a peak energy $E_{\mathrm
 peak}$, and a normalization. The results are summarized in
 Table~\ref{tab:spec}.

 To study the spectral evolution during the burst, we divide the
 spectra into three time intervals. As noticed from
 Figure~\ref{fig:lc}, there are two distinct peaks in the light curve
 in the 100--350 keV band at $T_0+$6.5 s and $T_0+$17.5
 s. Accordingly, the first, second, and last intervals are defined as
 between $T_0-$1.5--$T_0+$6.5 s (before the first peak),
 $T_0-$6.5--$T_0+$17.5 s (in-between the two peaks), and
 $T_0-$17.5--$T_0+$28.5 s (after the second peak), respectively. The
 fitting results of the time resolved spectra are also summarized in
 Table~\ref{tab:spec}.

\section{Discussions}

\begin{figure}[htbp]
  \begin{center}
   \rotatebox{0}{
    \FigureFile(80mm,80mm){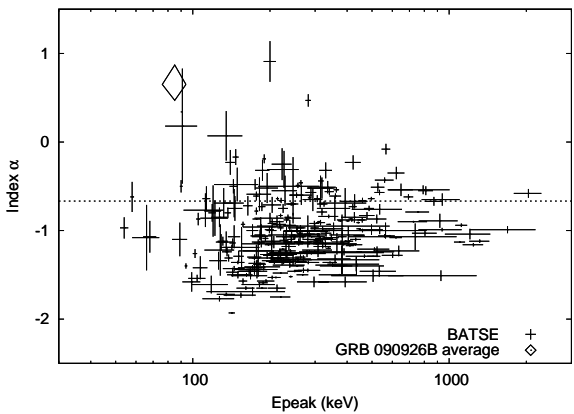}
    }
  \end{center}
  \caption{Scatter plot on the $E_{\mathrm peak}$-$\alpha$ plane.  
  The open diamond represents the time averaged spectrum of
  GRB\,090926B. 
  The BATSE sample from \citet{2006ApJS..166..298K} is plotted with 
  crosses.  
  }
  \label{fig:epalpha}
\end{figure}

 The earlier reports on the spectral analyses of GRB\,090926B
 from the Swift and Fermi teams suggested very flat spectra
 represented by a cut-off power law with a photon index 
 of $-0.52\pm0.24$ and $-0.13\pm0.06$ \citep{2009GCNR..246....1G,
 2009GCN..9957....1B}, and $E_{\mathrm peak}$ of 78.3$\pm$7.0 keV and 91$\pm$2
 keV, respectively.
 We confirm these results from the combined spectra of MAXI/GSC and Fermi/GBM. 
 Using the time averaged spectrum, we obtain the best fit cut-off power-law
 model with $\alpha = 0.44 \pm 0.14$ and 
 $E_{\mathrm peak}$ of $97\pm 7$ keV.
 The discrepancy between our result and that by the Fermi team mainly 
 comes from the difference of the time interval, rather than the energy 
 range, used for the spectral analyses; 
 while \citet{2009GCN..9957....1B} utilize the first 48.6 s data, we
 analyze only the first 30 s after the burst.  Indeed, from the analysis
 of GBM data alone, we had the result consistent with the result of
 joint fit analysis, when we limit the time range to first 30 s.
 When the ``GRB model'' is adopted, we obtain the index 
 $\alpha = 0.65 \pm 0.20$ and $E_{\mathrm peak} = 85 \pm 9$ keV.
 In either model, the obtained $\alpha$ value exceeds the line of death.
 Moreover, the spectral parameters are peculiar among GRBs.
 The $E_{\mathrm peak}$ and $\alpha$ values are plotted in the scatter plot in
 Figure~\ref{fig:epalpha} together with those of the BATSE sample taken from
 \cite{2006ApJS..166..298K}.
 The point of GRB\,090926B is apart from the ``main sequence'', and
 has both larger $\alpha$ and smaller $E_{\mathrm peak}$ compared with
 the majority of GRBs.

 Let us consider the mechanisms to produce such a high $\alpha$ spectrum.
 First, we investigate the possibility that the photons are heavily 
 absorbed somewhere between the source and the earth.
 To investigate this possibility, we fit the time averaged spectrum
 with an absorbed ``GRB model'', where the absorber is assumed to be located 
 at redshift of $z=1.24$ \citep{2009GCN..9947....1F}. 
 We obtain the best-fit absorption
 column density $N_{\mathrm H}$ consistent with zero%
 \footnote{The Galactic value of the absorption column density toward
 the burst direction is $1.91 \times 10^{20}$ cm$^{-2}$.} and $\alpha = 0.71$.
 Since the column density is often coupled with the power-law photon
 index, the confidence contours in the $\alpha$-N$_{\mathrm H}$ 
 space are drawn in Figure~\ref{fig:cont}.
 From this figure, a lower limit of $\alpha$ is found to be $0.42$
 (90\% confidence), and thus we conclude that the large photon index
 cannot be explained by an interstellar absorption.

 Then we have to consider the possibility that the burst has intrinsically
 large $\alpha$. \citet{2003A&A...406..879G} discussed various models 
 reproducing extremely hard spectra, including the bursts with the 
 low-energy photon indices $\alpha$ larger than $1$.  Similar discussions
 are presented in \citet{2005PASJ...57.1031S} about GRB\, 020813,
 which had the flattest spectrum among the bursts detected by HETE-2. 
 They studied the
 case of synchrotron self absorption and synchrotron self Compton
 as well. We can calculate the source radius and the electron number 
 density, from the redshift $z$=1.24 \citep{2009GCN..9947....1F},  
 $E_{\mathrm peak} = 85$ keV, and total energy of the burst $E_{\mathrm tot} =$
 4.3 $\times 10^{52}$ ergs, following \citet{2005PASJ...57.1031S}.

 For the case of synchrotron self absorption, we obtain the source radius 
 $r = 5.9 \times 10^{13}$ cm, minimum Lorentz factor of relativistic
 electrons $\Gamma_m = 400$, electron number density 
 $n = 10^{13}$ cm$^{-3}$, and magnetic field strength 
 $B = 2.9 \times 10^{5}$ gauss. As mentioned in 
 \citet{2005PASJ...57.1031S}, the peak flux calculated from the above
 parameters is inconsistent with the observed value unless we assume
 unusual physical conditions of the source.  For the case of synchrotron
 self Compton, we have $r = 3.1 \times 10^{16}$ cm, $\Gamma_m = 240$, 
 $n = 1.2 \times 10^{5}$ cm$^{-3}$, and $B = 24$ gauss. These results 
 are quite similar to the results of GRB\,020813, and then synchrotron self
 Compton model cannot be appropriate because of the large radius.
 Interestingly, the redshift of GRB\,020813 $z = 1.25$ 
 is close to that of GRB\,090926B ($z=1.24$) presumably.  The other 
 parameters also agree within an order of magnitude.
 Therefore, the discussion for GRB\,020813 is also 
 appropriate for GRB\,090926B.

 Jitter radiation, which is emitted by ultra-relativistic electrons in 
 highly nonuniform, small-scale magnetic fields, is studied as one of the 
 mechanisms responsible for such a hard low-energy index  
 \citep[e.g.][and references therein]{2009ApJ...702L..91M}. 
 \citet{2010ApJ...713..764R} 
 studied the observable spectral shape of the jitter radiation from
 various conditions of magnetic field.  According to their work, 
 despite the comprehensive search of the enormous parameter space, 
 it is not possible to find the condition to generate an index $\alpha$ 
 larger than 0. 
 On the other hand, we found from the fitting results that 
 the probability of $\alpha \leq 0$ is order of $10^{-6}$.
 Therefore, the spectrum of GRB\,090926B may not be
 produced by jitter radiation.

\begin{figure}[htbp]
  \begin{center}
   \rotatebox{0}{
    \FigureFile(80mm,80mm){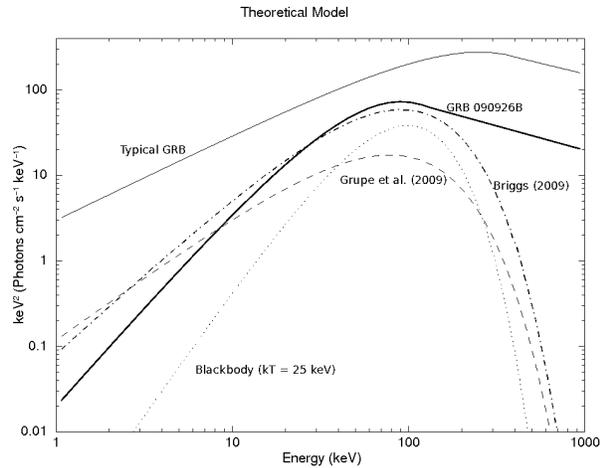}
    }
  \end{center}
  \caption{The best fit model of the time averaged spectrum is plotted with
  thick solid line in the $E F_{E}$ space. A typical
  GRB spectrum with $\alpha = -1$, $\beta = -2.5$, and $E_{\mathrm peak}$ = 
  250 keV is shown with thin solid line. A sample of blackbody spectrum 
  with kT = 25 keV is shown with thin dotted line.
  The spectral models reported to GCN by the Swift team 
  \citep{2009GCNR..246....1G} and the Fermi team 
  \citep{2009GCN..9957....1B} are plotted with dashed and dash-dotted lines
  respectively.
  Note that they represent the averaged spectra of the different 
  time interval in the burst (see text).}\label{fig:eemo}
\end{figure}

 Another possibility is that the spectrum is produced by thermal 
 radiation. Thermal components in the spectra of GRB prompt emission were 
 claimed to appear for several GRBs 
 \citep[and references therein]{2010ApJ...709L.172R}.  
 Some bursts showed spectra reproduced by blackbody $+$ power-law 
 components, and sometimes a power-law component was not necessary to 
 represent the observed spectra \citep{2006ApJ...652.1400R}. 
 Indeed, its spectrum 
 has a positive low-energy index and shows a narrow energy distribution.
 Figure~\ref{fig:eemo} shows the spectral energy distribution of this
 burst. Compared with a typical GRB (plotted with thin solid line),
 this burst has lower $E_{\mathrm peak}$ and a steeper rising part,
 which is rather close to the Rayleigh-Jeans part of a blackbody spectrum
 (plotted with thin dotted line).
 The thermal components of GRBs could be considered to be a contribution of
 the photosphere of GRB fireball.
 \cite{2000ApJ...530..292M} studied various cases of GRB spectra
 based on the internal shock model consisting of a photospheric component
 and Comptonized element by the pair plasma.

 Following \citet{1986ApJ...308L..43P}, the observable radius and 
 temperature of the blackbody radiation from the photosphere is constant 
 during the acceleration of the flow. Therefore, if we can estimate the 
 parameters of the blackbody, they are the parameters of the innermost part
 of the burst. In order to derive the photospheric radius and temperature,
 we test a model of simple blackbody radiation first. However,
 this model does not give a good fit owing to the tail-like component
 in the high energy part. This fact can be naturally interpreted as
 a temporal or a spatial superposition of the multiple temperature 
 rather than fully adiabatic and uniform photosphere.

 Then we adopted a model of Comptonized blackbody model
 \citep{1986PASJ...38..819N} for the purpose of fitting both the 
 blackbody-like component and high energy tail. 
 The results of this fit are shown in the bottom part of 
 Table~\ref{tab:spec}.

 \citet{2007ApJ...664L...1P} introduced the method to calculate 
 the parameters of photosphere from the observed spectral parameters,
 under the condition that the Lorentz factor of the plasma $\Gamma$
 is directly proportional to the radius of the fireball.
 There is a key parameter that indicates the ratio of the observed flux 
 and emitted flux
 \begin{eqnarray}
  {\cal R} &\equiv& \left( 
                \frac{F_{\mathrm BB}^{\mathrm ob}}{\sigma {T^{\mathrm ob}}^4}
                     \right)^{1/2} \, ,
 \end{eqnarray}
 where $\sigma$ is Stefan-Boltzmann constant, $F_{\mathrm BB}^{\mathrm ob}$ is
 the observed flux of the blackbody component, and $T^{\mathrm ob}$ is the 
 observed blackbody temperature.
 We used the equations 
 \begin{eqnarray}
  \Gamma &=& \left[ (1.06) (1+z)^{2} d_{\mathrm L} 
              \frac{Y F^{\mathrm ob} \sigma_{\mathrm T}}{ 2 m_{\mathrm p} c^3 {\cal R}}
               \right]^{1/4} \, ,
 \\
  r_0 &=& \frac{4^{3/2}}{(1.48)^{6}(1.06)^{4}}
           \frac{d_{\mathrm L}}{(1+z)^2} \left( 
            \frac{F_{\mathrm BB}^{\mathrm ob}}{Y F^{\mathrm ob}} 
             \right)^{2/3} {\cal R} \, ,
 \end{eqnarray}
 where $z$, $d_{\mathrm L}$, $\sigma_{\mathrm T}$, $m_{\mathrm p}$, and $c$ are the 
 redshift, luminosity distance to the source, Thomson cross section, 
 proton mass, and speed of light, respectively.  The parameter $Y$ is a 
 ratio between 
 the total fireball energy and the energy emitted in gamma-rays.
 The ratio of the total observed flux to the blackbody component
 $F^{\mathrm ob} / F_{\mathrm BB}^{\mathrm ob}$ depends on the
 energy range of integration, particularly on the upper bound, because
 Comptonized component dominates in higher energy part.
 For example, $F^{\mathrm ob} / F_{\mathrm BB}^{\mathrm ob} \sim 1.0$
 for the upper bound of 800 keV and 1.2 for 1200 keV.
 Consequently we introduce the renormalized parameter 
 $Y^{\prime} = Y F^{\mathrm ob} / F_{\mathrm BB}^{\mathrm ob}$ 
 instead of $Y$.
 The luminosity distance corresponding to the measured redshift
 $z=1.24$ is $d_{\mathrm L} = 2.68 \times 10^{28}$ cm under the standard 
 condition of $H_0 = 71$ km s$^{-1}$ Mpc$^{-1}$, 
 $\Omega_{\Lambda} = 0.73$, and $\Omega_{\mathrm M} = 0.27$.
 Using the temperature of the blackbody radiation $kT = 17.2$ keV
 and its observed flux $F_{\mathrm BB}^{\mathrm ob} = 3.0 \times 10^{-7}$
 erg cm$^{-2}$ s$^{-1}$, we find the Lorentz factor of the plasma 
 $\Gamma = (110 \pm 10) \,  Y^{\prime \, 1/4}$, and the physical size at the base of 
 the flow $r_0 = (4.3 \pm 0.9) \times 10^{9} \, Y^{\prime \, -3/2}$ cm. 
 These values are factor of a few different from the case of GRB\,970828, 
 GRB\,990510 \citep{2007ApJ...664L...1P}, or GRB\,090902B 
 \citep{2010ApJ...709L.172R}.

 According to \citet{2010arXiv1011.6005B}, the observed photospheric 
 spectrum is blackbody if the outflow energy is dominated by radiation
 rather than baryon up to the photospheric radius.  In other words,
 the fireball remained optically thick when the initial acceleration
 was saturated. This situation occurs in the case of low baryon load.
 The spectrum of GRB\,090926B may be one of the extreme example
 of the low baryon-load fireball.

\section{Conclusion}

 MAXI GSC observed the first 30 s of GRB\,090926B prompt emission.
 From the data of the scans before and after the burst, we could not
 find any signal of emissions with the flux limit of about 
 3 $\times 10^{-10}$ erg cm$^{-2}$ s$^{-1}$ (4--10 keV) for each scan.
 The joint spectral analysis with Fermi GBM reveals that the spectrum
 of GRB\,090926B shows a peculiar narrow shape. The spectral index
 $\alpha$ of time averaged spectrum is positive. The $E_{\mathrm peak}$ of 
 the burst is low relative to other bursts with such a hard spectral 
 indices. This hard spectral index cannot be realized by interstellar 
 absorption, synchrotron self absorption, synchrotron self Compton,
 nor jitter radiation.
 We find that the spectrum can be fit well by Comptonized blackbody
 model. The blackbody radiation can be interpreted as a photospheric
 emission of the GRB fireball.  Following the model by 
 \citet{2007ApJ...664L...1P}, we obtain the size of the base of the flow
 $r_0 = (4.3 \pm 0.9) \times 10^{9} \,  Y^{\prime \, -3/2}$ cm and Lorentz factor of the
 plasma $\Gamma = (110 \pm 10) \, Y^{\prime \, 1/4}$. 
 According to \citet{2010arXiv1011.6005B}, the observed photospheric 
 spectrum is blackbody if the outflow energy is dominated by radiation
 rather than baryon up to the photospheric radius. Therefore, the spectrum 
 of GRB\,090926B may be the example of the low baryon-load fireball.

\bigskip

This research was partially supported by the Ministry of Education, 
Culture, Sports, Science and Technology (MEXT), Grant-in-Aid No.19047001, 
20041008, 20244015, 20540230, 20540237, 21340043, 21740140, 22740120, and 
Global-COE from MEXT ``The Next Generation of Physics, Spun from 
Universality and Emergence'' and ``Nanoscience and Quantum Physics''.

\begin{figure}[htbp]
  \begin{center}
   \rotatebox{0}{
    \FigureFile(80mm,60mm){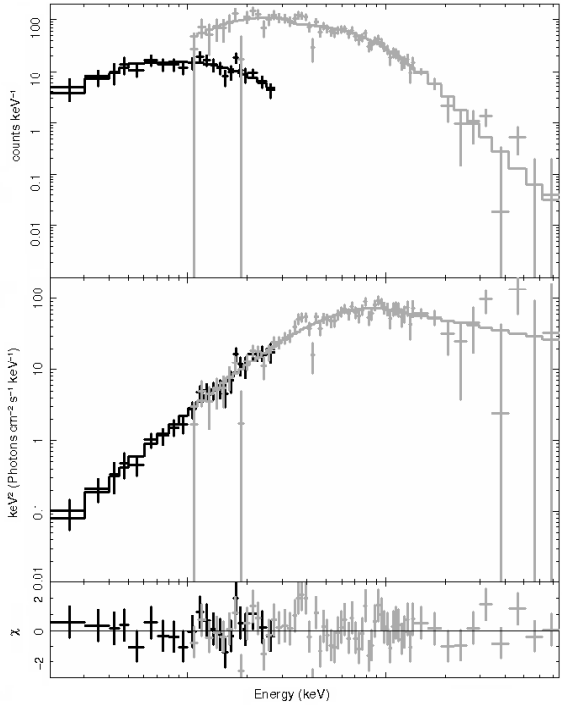}
    }
  \end{center}
  \caption{Time average spectrum of GRB\,090926B fitted with the
  ``GRB model''. The top panel shows the data and the folded model,
  and the middle panel shows the unfolded $E F_E$ spectrum.
  The residuals are plotted in the bottom panel.
  The data sets of the GSC and GBM are plotted in
  black and gray, respectively.
  }\label{fig:spec}
\end{figure}

\begin{figure}[htbp]
  \begin{center}
   \rotatebox{270}{
    \FigureFile(60mm,80mm){figure5.ps}
    }
  \end{center}
  \caption{Confidence contours of $\alpha$-$N_{\mathrm H}$ space for
  the time averaged spectrum.
  The confidence levels of 68\%, 90\%, and 99\% are shown.
  }\label{fig:cont}
\end{figure}

\begin{table*}[hp]
 \begin{center}
 \caption{Spectral parameters of GRB\,090926B}
 \label{tab:spec}
 \begin{tabular}{lcccc}
 \hline
            & \multicolumn{4}{c}{time interval} \\
  component & $T_0-$1.5 -- $T_0+$28.5 
              & $T_0-$1.5 -- $T_0-$6.5 
                & $T_0-$6.5  -- $T_0+$17.5 
                  & $T_0+$17.5  -- $T_0+$28.5  
  \\ \hline
  \multicolumn{5}{l}{ 
           {\bf cut-off power law}} \\
   $\chi^2$(DoF)  & 93.26 (83) 
                   & 20.56 (32) \footnotemark[$*$] 
                    & 65.41 (67) 
                     & 100.17 (81) \\
   Index  $\alpha$ (E$^{\alpha}$) & $0.44^{+0.14}_{-0.13}$ 
                   & $0.43^{+0.47}_{-0.33}$ 
                    & $0.76^{+0.31}_{-0.26}$ 
                     & $0.24^{+0.20}_{-0.17}$ \\
   $E_{\mathrm peak}$ (keV) & $97_{-6}^{+7}$ 
                   & $142_{-21}^{+30}$ 
                    & $76_{-7}^{+9}$ 
                     & $96_{-7}^{+8}$ \\
   Normalization
   \footnotemark[$\dagger$] & $4.3_{-0.3}^{+0.4}$  
                   & $1.6_{-0.5}^{+0.4}$ 
                    & $4.6_{-0.7}^{+0.8}$ 
                     & $6.9_{-0.8}^{+0.8}$ \\
  \multicolumn{5}{l}{ 
           {\bf GRB model}} \\
   $\chi^2$(DoF)  & 83.13 (82) \footnotemark[$*$] 
                   & 20.56 (31) 
                    & 61.90 (66)  
                     & 94.65 (80) \footnotemark[$*$] \\
   Index $\alpha$ (E$^{\alpha}$) & $0.65^{+0.22}_{-0.18}$ 
                   & $0.42^{+0.31}_{-0.31}$ 
                    & $1.07^{+0.61}_{-0.41}$ 
                     & $0.52^{+0.34}_{-0.25}$ \\
   Index $\beta$  (E$^{\beta}$) & $-2.51^{+0.29}_{-0.49}$ 
                   & -9.37 \footnotemark[$\ddagger$] 
                    & $-2.67^{+0.46}_{-1.71}$ 
                     & $-2.53^{+0.32}_{-0.71}$ \\
   $E_{\mathrm peak}$ (keV) & 85$_{-9}^{+9}$ 
                   & 143$_{-16}^{+29}$ 
                    & 67$_{-11}^{+12}$ 
                     & 83$_{-10}^{+11}$ \\
   Normalization
   \footnotemark[$\dagger$] & 4.5$_{-0.5}^{+0.6}$ 
                   & 1.7$_{-0.5}^{+0.5}$ 
                    & 5.2$_{-1.1}^{+2.0}$ 
                     & 7.2$_{-0.9}^{+1.1}$ \\
  \multicolumn{5}{l}{ 
           {\bf Comptonized blackbody} \footnotemark[$\S$]} \\
   $\chi^2$(DoF)  & 88.45 (83) 
                   & 22.31 (32) 
                    & 61.20 (67) \footnotemark[$*$] 
                     & 100.86 (81) \\
   temperature kT (keV) &  17.2$_{-1.0}^{+1.1}$ 
                    & 29.0$_{-4.5}^{+5.3}$ 
                     & 15.3$_{-1.5}^{+1.6}$ 
                      & 16.5$_{-1.3}^{+1.4}$ \\
   Normalization \footnotemark[$\|$] ($10^{10}$ cm)
                  & 4.9$_{-0.6}^{+0.7}$ 
                   & 1.6$_{-0.6}^{+0.8}$ 
                    & 5.0$_{-1.0}^{+1.2}$ 
                     & 6.1$_{-1.0}^{+1.2}$ \\
   optical depth $\tau$   
                  & 0.9$_{-0.2}^{+0.2}$ 
                   & 0.5$_{-0.5}^{+0.6}$ 
                    & 0.7$_{-0.3}^{+0.3}$ 
                     & 0.8$_{-0.2}^{+0.2}$ \\
  \hline
    \multicolumn{5}{@{}l@{}}{\hbox to 0pt{\parbox{180mm}{\footnotesize
        \footnotemark[$*$] Best fit model to the spectrum.
        \par\noindent
        \footnotemark[$\dagger$] Normalizations are in the unit of 
         $10^{-2}$ photons cm$^{-2}$ s$^{-1}$ keV$^{-1}$ at 15 keV.
        \par\noindent
        \footnotemark[$\ddagger$] Errors are not available.
        \par\noindent
        \footnotemark[$\S$] The electron temperature is fixed to 50 keV.
        \par\noindent
        \footnotemark[$\|$] Normalizations are given as a radius of 
         blackbody, on the assumption of redshift $z$=1.24.
      }\hss}}
 
 \end{tabular}
 \end{center}
\end{table*}


\begin{thebibliography}{20}
\bibitem[Band et al.(1993)]{1993ApJ...413..281B} Band, D., et al.\ 1993, 
\apj, 413, 281 

\bibitem[Beloborodov(2010a)]{2010MNRAS.407.1033B} Beloborodov, A.~M.\ 2010a, 
\mnras, 407, 1033 

\bibitem[Beloborodov(2010b)]{2010arXiv1011.6005B} Beloborodov, A.~M.\ 2010b, 
arXiv:1011.6005

\bibitem[Briggs(2009)]{2009GCN..9957....1B} Briggs, M.~S.\ 2009, 
GRB Coordinates Network, 9957, 1 

\bibitem[Epstein 
\& Petrosian(1973)]{1973ApJ...183..611E} Epstein, R.~I., 
\& Petrosian, V.\ 1973, \apj, 183, 611 

\bibitem[Fynbo et al.(2009)]{2009GCN..9947....1F} Fynbo, J.~P.~U., et al.\ 
2009, GRB Coordinates Network, 9947, 1 

\bibitem[Ghirlanda et al.(2003)]{2003A&A...406..879G} Ghirlanda, G., 
et al.\ 2003, \aap, 406, 879 

\bibitem[Grupe et al.(2009)]{2009GCNR..246....1G} Grupe, D., et al.\ 2009, 
GCN Report, 246, 1 

\bibitem[Kaneko et al.(2006)]{2006ApJS..166..298K} Kaneko, Y., et al.\ 
2006, \apjs, 166, 298 

\bibitem[Lazzati \& Begelman(2010)]{2010ApJ...725.1137L} 
Lazzati, D., \& Begelman, M.~C.\ 2010, \apj, 725, 1137 

\bibitem[Matsuoka et al.(2009)]{2009PASJ...61..999M} Matsuoka, M., et al.\ 
2009, \pasj, 61, 999 

\bibitem[Medvedev(2000)]{2000ApJ...540..704M} Medvedev, M.~V.\ 2000, 
\apj, 540, 704

\bibitem[Medvedev et al.(2009)]{2009ApJ...702L..91M} Medvedev, M.~V., 
Pothapragada, S.~S., \& Reynolds, S.~J.\ 2009, \apjl, 702, L91 

\bibitem[M{\'e}sz{\'a}ros \& Rees(2000)]{2000ApJ...530..292M} 
M{\'e}sz{\'a}ros, P., \& Rees, M.~J.\ 2000, \apj, 530, 292 

\bibitem[Mihara et al.(2011)]{mihara.gsc}
Mihara, T., et al.\ 2011, \pasj, accepted, arXiv:1103.4224

\bibitem[Morii et al.(2009)]{2009GCN..9943....1M} Morii, M., et al.\ 2009, 
GRB Coordinates Network, 9943, 1 

\bibitem[Nishimura et al.(1986)]{1986PASJ...38..819N} Nishimura, J., 
Mitsuda, K., \& Itoh, M.\ 1986, \pasj, 38, 819

\bibitem[Paczynski(1986)]{1986ApJ...308L..43P} Paczynski, B.\ 1986, \apjl, 
308, L43 

\bibitem[Pe'er et al.(2007)]{2007ApJ...664L...1P} Pe'er, A., Ryde, F., 
Wijers, R.~A.~M.~J., M{\'e}sz{\'a}ros, P., 
\& Rees, M.~J.\ 2007, \apjl, 664, L1 

\bibitem[Preece et al.(2002)]{2002ApJ...581.1248P} Preece, R.~D., et al.\ 
2002, \apj, 581, 1248 

\bibitem[Rees \& Meszaros(1994)]{1994ApJ...430L..93R} Rees, M.~J., \& 
Meszaros, P.\ 1994, \apjl, 430, L93 

\bibitem[Reynolds et al.(2010)]{2010ApJ...713..764R} Reynolds, S.~J., 
Pothapragada, S., \& Medvedev, M.~V.\ 2010, \apj, 713, 764 

\bibitem[Ryde(2004)]{2004ApJ...614..827R} Ryde, F.\ 2004, \apj, 614, 827 

\bibitem[Ryde et al.(2006)]{2006ApJ...652.1400R} Ryde, F., et al.\ 2006, 
\apj, 652, 1400 

\bibitem[Ryde et al.(2010)]{2010ApJ...709L.172R} Ryde, F., et al.\ 2010, 
\apjl, 709, L172 

\bibitem[Sato et al.(2005)]{2005PASJ...57.1031S} Sato, R., et al.\ 2005, 
\pasj, 57, 1031 

\bibitem[Sugizaki et al.(2011)]{2011arXiv1102.0891S} Sugizaki, M., et al.\ 
2011, \pasj, accepted, arXiv:1102.0891 

\bibitem[Tomida et al.(2011)]{2011PASJ...63..397T} Tomida, H., et al.\ 
2011, \pasj, 63, 397 

\end{thebibliography}
\end{document}